\documentclass{WileyMSP-template}

\usepackage{amsmath}
\usepackage{amssymb}
\usepackage{graphicx}
\usepackage{textcomp}
\usepackage{gensymb}
\usepackage{grffile}
\usepackage{etoolbox}
\usepackage{gensymb}
\usepackage{graphicx}
\usepackage{subfigure}
\usepackage{amsmath}
\usepackage{amssymb}
\usepackage{dsfont}
\usepackage{hyperref}
\usepackage{multirow}
\usepackage{rotating}
\usepackage{epstopdf}
\usepackage{color}
\usepackage{wrapfig}
\usepackage{fancyhdr}
\usepackage{lipsum}
\usepackage{tabularx}
\DeclareUnicodeCharacter{0096}{_}
\DeclareUnicodeCharacter{2009}{ }
\DeclareUnicodeCharacter{2212}{-}
\DeclareUnicodeCharacter{03B4}{$\delta$}

\begin{document}

\pagestyle{fancy}

\title{Nonlinear coherent light-matter interaction in 2D MoSe$_2$ nanoflakes for all-optical switching and logic applications}

\maketitle



\author{Sk Kalimuddin$^{\dagger}$, Biswajit Das$^{\dagger}$, Nabamita Chakraborty, Madhupriya Samanta,  Satyabrata Bera, Arnab Bera, Deep Singha Roy, Suman Kalyan Pradhan, Kalyan K. Chattopadhyay* and Mintu Mondal* }\\
\begin{affiliations}
	Sk Kalimuddin, Dr. B. Das, S. Bera, A. Bera, D. S. Roy, Dr. S. K. Pradhan, Dr. M. Mondal\\
	School of Physical Sciences, Indian Association for the Cultivation of Science, Jadavpur, Kolkata 700032, India\\
	Email Address:Mintu.Mondal@iacs.res.in\\
	M. Samanta\\
	Department of Electronics \& Telecommunication Engineering, Jadavpur University, Kolkata-700032, India\\
	N. Chakraborty, Prof. K. K. Chattopadhyay\\
	Thin film \& Nanoscience lab, Department of Physics, Jadavpur University, Kolkata-700032, India\\
	Email Address:kalyan\_chatopadhyay@yahoo.com\\
	$\dagger$ These authors contributed equally to this work\\
\end{affiliations}



\begin{abstract}
We report a strong nonlinear optical response of 2D MoSe$_2$ nanoflakes (NFs) through spatial self-phase modulation (SSPM) and cross-phase modulation (XPM) induced by nonlocal coherent light-matter interactions. The coherent interaction of light and MoSe$_2$ NFs creates the SSPM of laser beams, forming concentric diffraction rings. The nonlinear refractive index ($n_2$) and the third-order broadband nonlinear optical susceptibility ($\chi^{(3)}$) of MoSe$_2$ NFs are determined from the self diffraction pattern at different exciting wavelengths of 405, 532, and 671 nm with varying the laser intensity. The evolution and deformation of diffraction ring patterns are observed and analyzed by the `wind-chime' model and thermal effect. By taking advantage of the reverse saturated absorption of 2D SnS$_2$ NFs compared to MoSe$_2$, an all-optical diode has been designed with MoSe$_2$/SnS$_2$ hybrid structure to demonstrate the nonreciprocal light propagation. Also a few other optical devices based on MoSe$_2$ and other semiconducting materials such as Bi$_2$Se$_3$, CuPc, and graphene have been investigated. The all-optical logic gates and all-optical information conversion have been demonstrated through the XPM technique using two laser beams. The proposed optical scheme based on MoSe$_2$ NFs has been demonstrated as a potential candidate for all-optical nonlinear photonic devices such as all-optical diodes and all-optical switches.

\end{abstract}

\section{Introduction}
Optical responses of two-dimensional (2D) nanomaterials have drawn much attention due to their strong nonlinear characteristics and broadband Kerr nonlinearity for their potential applications in photonic devices. The strong light-matter interactions in two-dimensional (2D) materials produce many interesting nonlinear optical phenomena, and quasiparticle excitations that can have far-reaching implications in science and technology \cite{Liao2020a, Wu2020a, Zhang2019}. Therefore, 2D nanomaterials have attracted significant attention from fundamental physics to applied sciences as a scientific platform for studying nonlinear optical phenomena and optoelectronic devices \cite{You2019, Tan2020}. The nonlinear optical response has been investigated using the following three main methods that are four-wave mixing, Z-scan, and spatial self-phase modulation (SSPM) spectroscopy \cite{Turitsyn2015, Wang2021, Wu2016}. The four-wave mixing and Z-scan spectroscopy require a complicated experimental setup, while similar results can be obtained in the SSPM with a simple, straightforward technique. Therefore, SSPM spectroscopy has become an effective tool for studying the nonlinear optical properties of the materials like nonlinear refractive index (n$_2$), and third-order nonlinear susceptibility ($\chi^{(3)}$) \cite{Shen1984, Boyd2020, Jia2019a}. In this experiment, strong coherent light-matter interaction creates spatial phase modulation of the incident laser light and produces diffraction ring patterns in the far field. Durbin et al. first observed the SSPM in liquid crystals, and recently, in 2011, Wu et al. investigated the third-order nonlinear susceptibility in exfoliated graphene by the SSPM method \cite{Durbin1981, Durbin1982, Wu2011}. After that, there has been a significant increase in interest in SSPM, and many groups have studied the nonlinear response of different types of materials. 

The nonlinear interaction of light and materials leads to self diffraction rings/patterns in the far-field due to the SSPM. The formation, expansion, and collapse process of the diffraction rings with time has been described by the `wind-chime' model based on the nonlocal electric coherent theory \cite{Wu2015}. In this model, the samples are being polarized because of interaction between the samples (layered nanostructures) and incident coherent light beam \cite{Xiao2020, Zhang2016}. Based on the energy relaxation process, the samples are reoriented from an arbitrary angle to the direction of the electric field of the incident light beam \cite{He2022}. Although the above model describes the SSPM phenomena quite well, there are still a few open questions, like how the relaxation process occurs with incident light energy, how the relaxation time depends upon the solvent viscosity, and the role of polar and non-polar solvents? Therefore a detailed study of SSPM with different materials and solvents is desired for in-depth understanding.

Recently, a significant amount of work has been done on various 2D materials, such as graphene, topological insulators, perovskite, transition metal dichalcogenides (TMDs), MXenes, black phosphorus, etc. \cite{Wu2019, Shan2019, Jia2018, Li2020, Miao2017}, and different hybrid structures are also studied for photonic device application based on the SSPM method \cite{Wu2020, Li2016, Sadrolhosseini2016}. The transition metal dichalcogenides (TMDs) are attractive candidates due to their unique tunable bandstructure and superior optic-electronic properties. TMDs are composed of transition metal elements intercalated between chalcogen elements in layered structures\cite{Liao2019}. In these materials, the adjacent layers are held together by the weak van der Waals forces \cite{Li2020}. One of the most prevalent members of TMD's family is MoSe$_2$ which can be easily synthesized using various methods, including hydrothermal, sonochemical, chemical vapor deposition, etc. \cite{Maity2020, Tang2016, Kristl2003, Etzkorn2005}. The  MoSe$_2$ are being widely explored for multiple potential applications, such as water splitting, catalysts, batteries, photoelectrochemical, solar cells, sensors, etc. \cite{Yang2017, Mao2015, Niu2017, Kline1981, Li2015, Pataniya2021}. However, the applications of MoSe$_2$ in photonics are still in their infancy \cite{Wang2016}.  

So far, the SSPM method has been used to determine the third-order nonlinear susceptibility and the nonlinear refractive index of various semiconducting materials, but comparatively few optical devices are made for practical applications \cite{Liao2019, Shi2015}. Very few groups have devoted their attention to making novel optical logic gates using the SSPM technique for photonic devices \cite{Wu2020, Song2020,Liao2020}. These all-optical logic gates can open the paths toward many other all-optical signal processing such as optical networking, optical switching, optical computing, optical transmission, etc.\cite{Wu2019}. Recently, all-optical switching and information converters based on SSPM have been demonstrated \cite{Song2020}. However, there are no reports/works on photonic devices based on MoSe$_2$ for all-optical logic applications. 
 
In this work, we have experimentally investigated the nonlinear optical (NLO) properties of 2D MoSe$_2$ NFs using SSPM spectroscopy and designed a novel photonic diode. Furthermore, we have also presented a hybrid structure-based optical device for the realization of the `OR' function for all-optical logic gates. We believe that this study will enhance the understanding of the nonlinear optical process in 2D materials and can lead to the development of optical devices based on the self-diffraction of light.

\section{Results and discussion}
\begin{figure} [t]
	\centering
	\includegraphics[width=0.75\columnwidth]{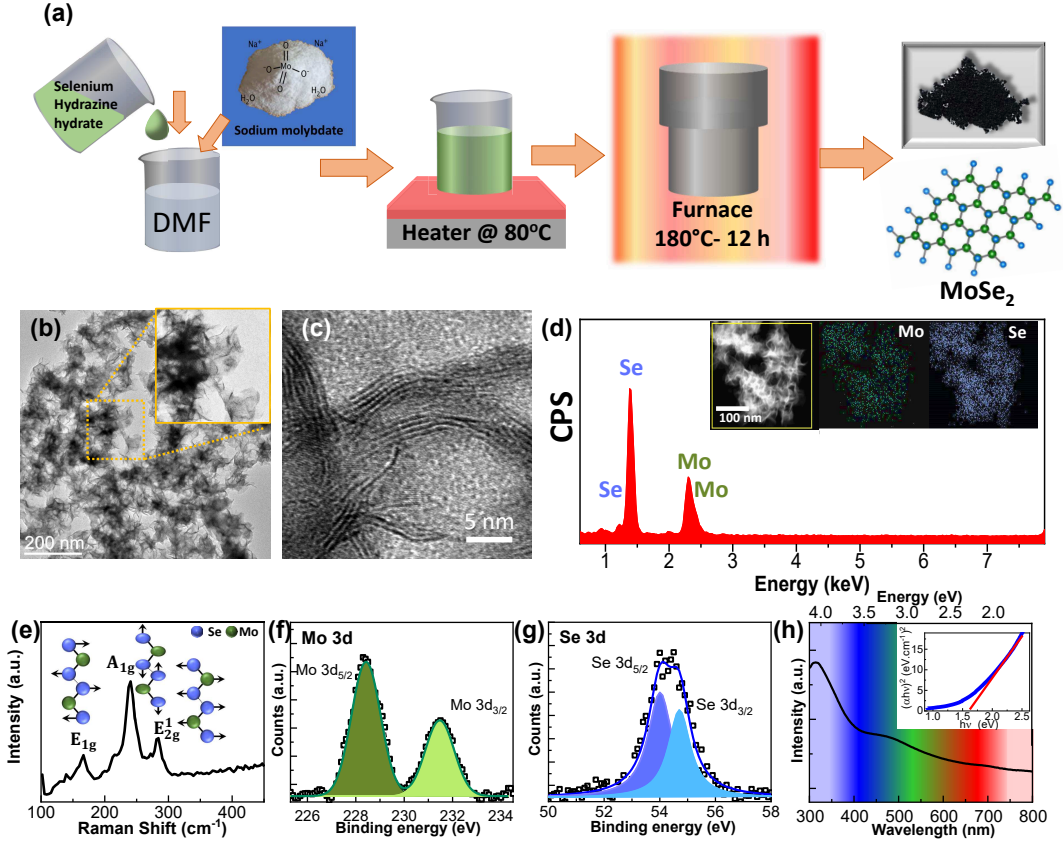}
	\caption{\textbf{Sample preparation and basic characterizations.} (a) Schematic presentation of the synthesis of MoSe$_2$ NFs. (b) low and high magnification (inset) TEM image of the MoSe$_2$ NFs. (c) HRTEM image of the individual NFs with few layer thickness. (d) EDS spectrum of the MoSe$_2$ NFs with their elemental mapping (inset) Mo, and Se, respectively. (e) Raman spectrum of the MoSe$_2$ NFs. High resolution (f) Mo 3d and (g) Se 3d binding energy spectrum of MoSe$_2$ NFs. (h) UV-Vis absorption spectrum of MoSe$_2$ NFs and its Tauc plot (inset). }
	\label{fig 1:1}
\end{figure}

\subsection{Nonlinear Kerr effect : Nonlinear refractive index of MoSe$_2$ nanoflakes}

\begin{figure}[t]
	\centering
	\includegraphics[width=0.75\columnwidth]{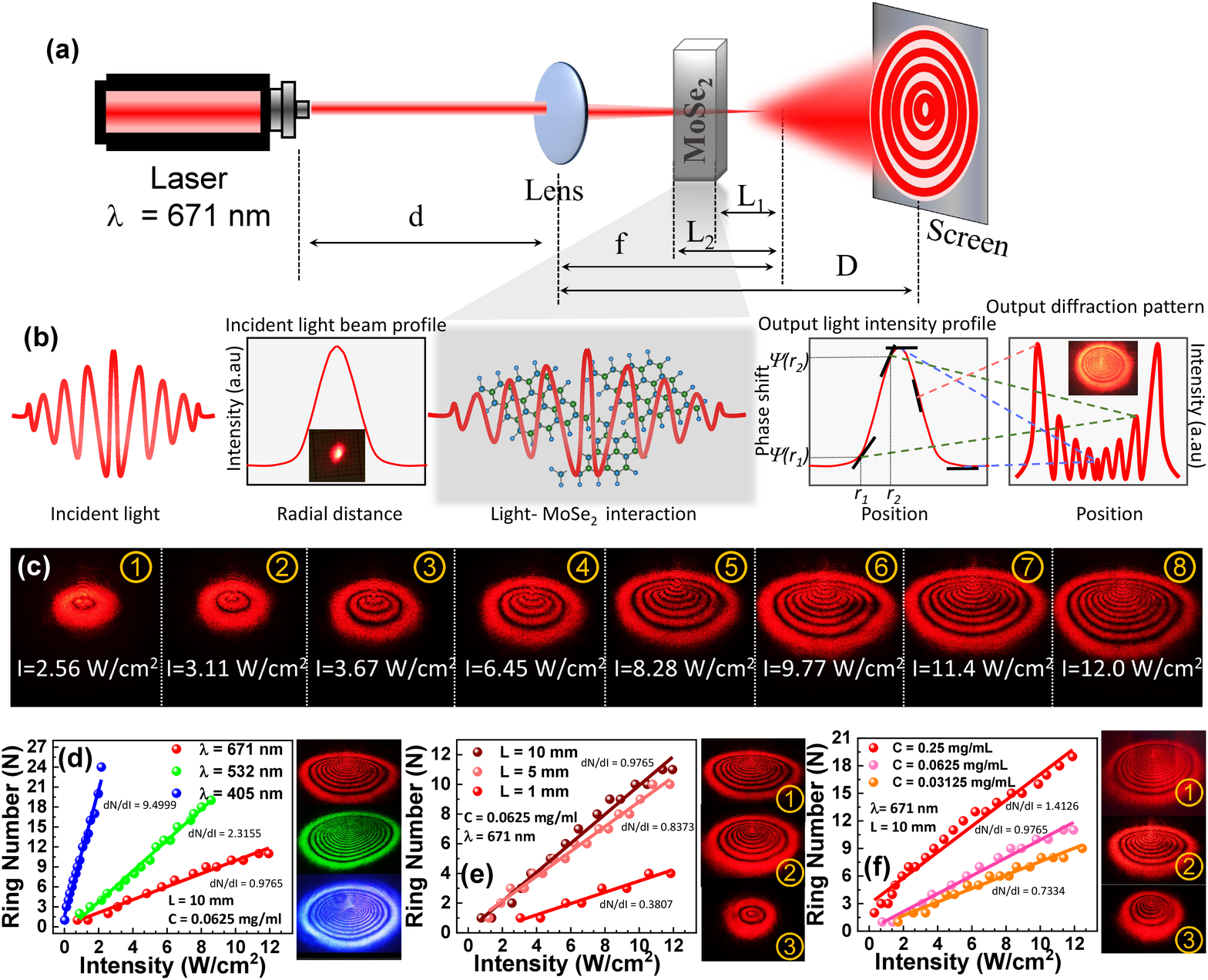}
	\caption{\textbf{Schematic of the experimental setup and SSPM in MoSe$_2$ NFs.} (a) Experimental setup for SSPM study of MoSe$_2$ NFs. (b) Schematic of the microscopic light-matter interaction process:  The diffraction ring produced through the coherent interaction of light and suspended MoSe$_2$ NFs. (c) The evolution of diffraction ring patterns on the screen  with the intensity of incident laser ($\lambda$= 671 nm). (d) The plot of diffraction ring numbers vs laser intensity for $\lambda$ = 405, 532 and 671 nm. (e) The variation of diffraction ring numbers with incident laser ($\lambda$= 671 nm) intensity with cuvette lengths of 1, 5 and 10 mm. (f) The diffraction ring number vs intensity of laser ($\lambda$ =671 nm) with various concentration of  MoSe$_2$  (0.25 mg/ml, 0.0625 mg/ml, and 0.03125 mg/ml).}
	\label{fig 2:2}
\end{figure}

The preparation of MoSe$_2$ NFs and its characterization are discussed in the experimental section \ref{sec:Experimental} and presented in Figure {\ref{fig 1:1}}. The Kerr nonlinear effect plays an important role in the study of the nonlinear optical response of 2D semiconducting (MoSe$_2$) materials. According to the nonlinear Kerr effect, the nonlinear refractive index can be expressed as $n = n_0 + n_2I$,  where $I$ is the incident light intensity, $n_0$ and $n_2$ are the linear and nonlinear refractive index coefficient of the material, respectively \cite{Wang2014}. The MoSe$_2$ is expected to have a strong Kerr nonlinear response. Therefore, the phase has a nonlinear modulating behavior to the transverse intensity profile of the incident Gaussian laser beam, and this nonlinear phase shift($\Delta\psi(r)$) can be expressed by \cite{Jia2018},

\begin{equation}
	\Delta\psi (r) = \frac{2\pi n_0}{\lambda} \int_{0}^{L_{eff}}n_2I(r,z)dz
	\label{eq:phase}
\end{equation}
where $\lambda$ is the wavelength of the laser, $r \in [0,+\infty)$ is the laser radial position, $L_{eff}$ is the effective transmission length of the laser passing through the cuvette.  This `self' phase shift ($\Delta\psi$) gets modulated due to the nonlinear Kerr effect through the change in optical refractive index, $n$, as function intensity of the coherent Gaussian laser beam. Therefore,  the propagating beam forms self-diffraction patterns in the far-field due to the spatial self-phase modulation (SSPM). The schematic representation of SSPM due to the light-matter interaction between the incident laser beam and the reoriented MoSe$_2$ NFs is shown in \ref{fig 2:2} (b). There are at least two different points $r_1$ and $r_2$ in the resultant outgoing Gaussian light, where the slopes of the distribution curve are same ($d\Delta\psi / dr)_{r=r_1}$~=~($d\Delta\psi / dr)_{r=r_2}$ and they have same phase. Therefore, the output light intensity profile with the same slope points maintains constant phase differences. The diffraction ring patterns are either bright or dark as given by the following condition \cite{Wu2011},

\begin{equation}
	\Delta\psi(r_1)-\Delta\psi(r_2)=2M\pi
	\label{eq:phase1}
\end{equation}

where $M$ is the integer number. The odd and even values of the $M$ correspond to the dark and bright diffraction ring, respectively. Also the path difference, $L_{eff}$ of the laser beam inside the cuvette can be determined from the following equation {\ref{eq:leff}} \cite{Zhang2016, Xiao2021}

\begin{equation}
	L_{eff} = \int_{L_1}^{L_2}\Bigg(1+\frac{z^2}{z^2_0}\Bigg)^{-1}dz = z_0tan^{-1}\left[\frac{z}{z_0}\right]_{L_1}^{L_2};   z_0 = \frac{\pi \omega_0^2}{\lambda}
	\label{eq:leff}
\end{equation}

where $L_1$ and $L_2$ are the distance from the focus ($f$) to the side of the quartz cuvette. The central intensity profile of the transmitted Gaussian beam can be expressed as $I(0,z) = 2I$, where $I$ is the average intensity of the incident laser, $z_0$ is the diffraction length and $w_0$ is the 1/$e^2$ beam radius.  From the equation {\ref{eq:phase}} and {\ref{eq:leff}}, the nonlinear refractive index ($n_2$) of MoSe$_2$ NFs can be determined as

\begin{equation}
	n_2 = \frac{\lambda}{2n_0L_{eff}}\frac{dN}{dI}
	\label{eq:n2final}
\end{equation}

Hence, the third-order nonlinear susceptibility,($\chi^{(3)}_{total}$), can be determined for  MoSe$_2$ NFs using the NLO properties as \cite{Boyd2020, Wu2019, Shan2019b}: 

\begin{equation}
	\chi^{(3)}_{total} = \frac{cn_0^2}{12\pi^2} 10^{-7}n_2 ~~ (esu)
	\label{eq:chi_total}
\end{equation}

where $c$ is the velocity of light in the free space. However, the available number of active 2D NFs of MoSe$_2$ present in the solvent has put a significant role on ${\chi^{(3)}_{total}}$, so it is necessary to determine the third-order nonlinear susceptibility for a single-layer (${\chi^{(3)}_{monolayer}}$). Hence, the ${\chi^{(3)}_{monolayer}}$ for the single-layer MoSe$_2$ can be determined by \cite{Shan2019b}

\begin{equation}
	\chi^{(3)}_{total} = \chi^{(3)}_{monolayer} \times N^2_{eff}
	\label{eq:chi_mono}
\end{equation}

where, $N_{eff}$ represents the effective number of layers of material. This  ${\chi^{(3)}}$ has a strong dependence on the electric field strength of the incident laser beam. The detailed calculation has been provided in the appendix section. The $\frac{dN}{dI}$ is a crucial parameter for the determination of the NLO refractive index of 2D materials.\\

The Figure {\ref{fig 2:2}}(c)(\textcircled{1}-\textcircled{8}) shows the diffraction ring patterns projected on the screen as a function of intensity of the incident light ($\lambda = 671\,nm$). With the increasing intensity of the incident light, the horizontal diameter of the rings and also the ring numbers increase linearly. The same experiment has been repeated for the other two lasers with $\lambda$ = 532 nm and 405 nm, and the results are presented in Figure {\ref{fig 2:2}}(d). Figure {\ref{fig 2:2}}(d) shows the variation of the diffraction ring numbers as a function of the intensity of the laser beams. From the linear fitting of the corresponding data, the estimated slopes ($\frac{dN}{dI}$) are 0.9765, 2.3155 and 9.4999 for $\lambda$ = 671, 532 and 405 nm, respectively. The results reveal that $\frac{dN}{dI}$ increases with decreasing wavelength of the laser lights. Here, the larger photon energy gives rise to higher SSPM effects consistent with other previously reported results \cite{Shan2019, Wu2019}. NLO responses of 2D materials are dependent not only on the photonic energy but also on a few other parameters, like available active materials and the effective length travel by the laser inside the medium. To study this effect, the concentration of MoSe$_2$ and the cuvette thickness have varied. Figure {\ref{fig 2:2}}(e) shows the variation of the diffraction ring numbers as a function of laser light intensity with $\lambda$=671 nm for three different cuvette thicknesses. The number of diffraction rings has a dependence on the thickness of the cuvette, and their slopes are found to be 0.9765, 0.8373, and 0.3807 for cuvette length $L$ = 10, 5, and 1 mm, respectively. The corresponding diffraction ring patterns at highest intensity are presented in Figure {\ref{fig 2:2}(e)(\textcircled{1}-\textcircled{3}). Hence, the available light-matter interaction due to the presence of active materials inside the medium is also studied, as a function of laser light intensity with $\lambda$=671 nm, by varying the concentration of MoSe$_2$ NFs with the fixed cuvette thickness ($L$=10 mm). The Figure {\ref{fig 2:2}(f)} shows the variations of diffraction rings with the incident light intensity for different concentrations, $C$ = 0.125, 0.0625 and 0.03125 mg/ml and the corresponding slopes are 1.4126, 0.9760, and 0.7334, respectively. The diffraction images for the highest intensity are shown in Figure {\ref{fig 2:2}(f)(\textcircled{1}-\textcircled{3})} for $C$ = 0.25, 0.0625 and 0.03125 mg/ml, respectively. From the above experiments, the n$_2$ and $\chi^{(3)}$ are calculated and presented in the Table {\ref{tab:1}}. As evident from the results, the increment of n$_2$ and $\chi^{(3)}$ occurs due to an increase in the energy of the incident laser beam and the active material concentration. 
These results reveal that both n$_2$ and $\chi^{(3)}$ increase with the increasing energy of the incident laser beam. Whereas with reducing the effective travel length of the laser beam inside the medium by changing the thickness of the cuvette, n$_2$ and $\chi^{(3)}$ values are also increased. Similarly, with increasing the concentration of the active material, the n$_2$ and $\chi^{(3)}$ values are also increased.
		
	\begin{table}
		\centering
		\caption{Nonlinear refractive index and nonlinear susceptibility estimated using SSPM spectroscopy}
		\label{tab:1}
		\begin{tabular}{cccccccc}
			\hline
			&$\lambda$ &$L$  &$C$ &dN/dI &n$_2$ &$\chi^{(3)}_{total}$ &$\chi^{(3)}_{monolayer}$\\
			&[nm] &[mm] &[mg/ml] &[cm$^2$/W] &[cm$^2$/W] &[esu] &[esu]\\
			\hline
			&671 &10 & 0.0625 & 0.975652 &6.6 x 10$^{-6}$ &0.00428 &1.35 x 10$^{-8}$\\
			&532 &10 & 0.0625 & 2.3155 &1.3 x 10$^{-5}$ &0.00706 &2.23 x 10$^{-8}$\\
			&405 &10 & 0.0625 & 9.49993 &6.1 x 10$^{-5}$ &0.01537 &4.86 x 10$^{-8}$\\
			&671 &5 & 0.0625 & 0.83735 &1.1 x 10$^{-5}$ &0.00735 &2.32 x 10$^{-8}$\\
			&671 &1 & 0.0625 & 0.38072 &2.6 x 10$^{-5}$ &0.01672 &5.29 x 10$^{-8}$\\
			&671 &10 & 0.25 & 1.41262 &9.6 x 10$^{-6}$ &0.0062 &1.96 x 10$^{-8}$\\
			&671 &10 & 0.03125 & 0.73344 &4.9 x 10$^{-6}$ &0.00322 &1.01 x 10$^{-8}$\\
			\hline
		\end{tabular}
	\end{table}%
	
Here, we have observed that the number of diffraction rings increases with increasing cuvette thickness and the concentration of MoSe$_2$ NFs. The above results can be understood from the interaction of light with the suspended NFs. The variation of  $\chi^{(3)}$ and n$_2$ are similar in nature, as suggested from the equation \ref{eq:chi_total}, whereas n$_2$ has strong dependence on $\lambda$ of incident laser, effective transmission length of the laser propagating through the medium and $\frac{dN}{dI}$ generates from the light-matter interaction. The electrons of the NFs coherently oscillate due to the strong light-matter interaction. With the increasing cuvette thickness or NFs density in the solution, effective interaction of the laser beam with materials increases. These results are observed due to the larger spatial phase shift of the laser beam leading to creates more diffraction rings. 

The obtained n$_2$ and $\chi^{(3)}$ for MoSe$_2$ NFs  are found to be comparatively larger than other family members of transition metal dichalcogenides (TMDs) (see Table~\ref{tab:1} and section S2 of supplementary). The larger NLO responses compared to other TMDs can be attributed to the morphological changes and large surface area of MoSe$_2$\cite{Liu2019}.

\begin{figure} [t]
	\centering
	\includegraphics[width=0.75\columnwidth]{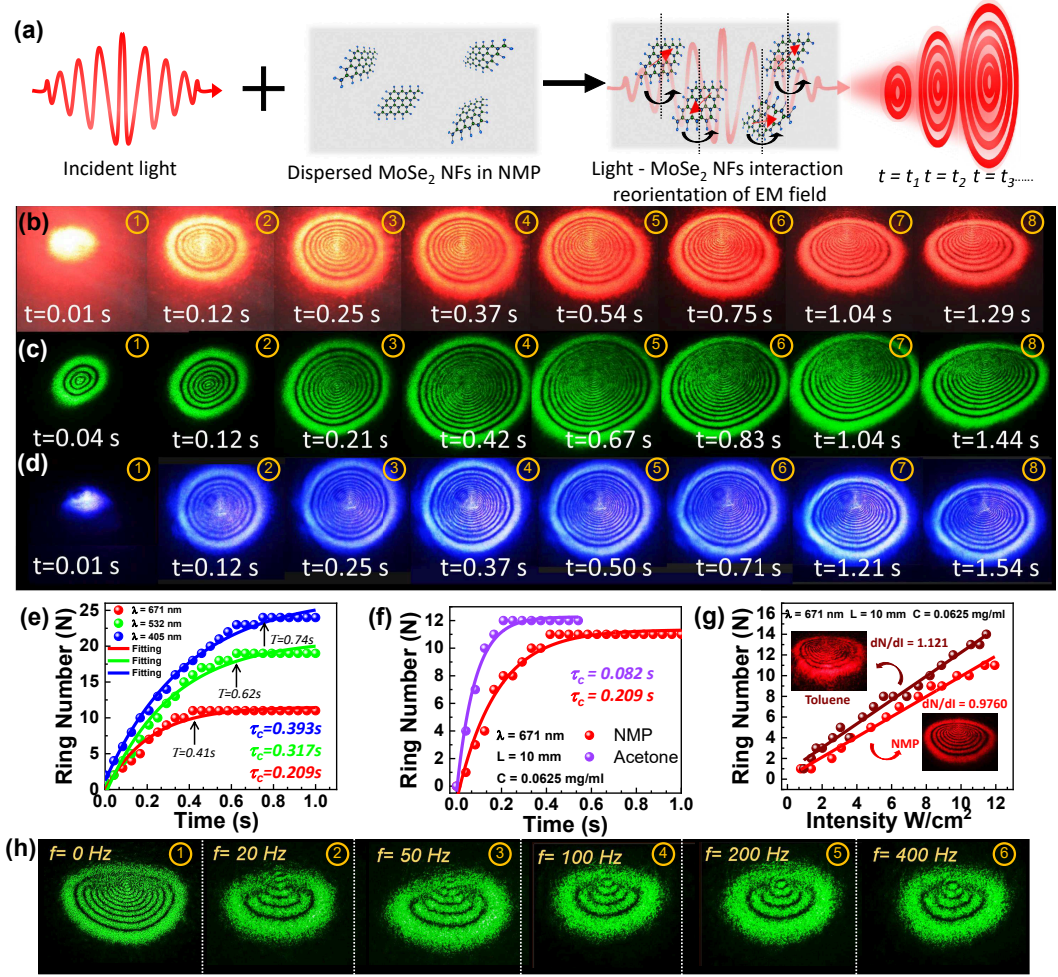}
	\caption{\textbf{The emergence of coherent light-matter interaction (`Wind-chime' model) and the thermal lens effect.} (a) Schematic presentation of the formation of SSPM patterns for MoSe$_2$ NFs via wind-chime model. (b-d) The diffraction ring pattern formed on the screen as function of time for $\lambda$= 671, 532, and 405 nm, respectively. (e) The evolution of the diffraction ring numbers with time at highest intensity for $\lambda$ = 671, 532, and 405 nm. (f) The plot of diffraction ring numbers with time in different viscous medium (NMP and acetone) for $\lambda$ = 671 nm. (g) The diffraction ring numbers through SSPM in both polar (NMP) and non-polar (toluene) solvent at $\lambda$= 671 nm. (h) SSPM patterns at different mechanical chopper modulation frequencies (0~-~400~Hz). }
	\label{fig 3:3}
\end{figure}

\subsection{SSPM: Mechanism of light-matter interaction}
In the above-mentioned nonlinear Kerr effect leading to SSPM, the $n_2$ and $\chi^{(3)}$ are determined for the suspended MoSe$_2$ NFs. The self-phase modulated diffraction ring numbers increase with time to a maximum value. The time required to achieve that maximum diameter at a fixed intensity is described by the `wind-chime' model proposed by Wu et al. \cite{Wu2015}. According to this model, the intense laser beam polarizes (or reorients) the suspended NFs under the influence of the optical electric field through the energy relaxation process. As a result, the number of diffraction rings increases with time due to the enhanced phase modulation of the Gaussian beam from coherent interaction with the reoriented NFs.  Here, the suspended NFs are considered individual domains and the reoriented NFs interact coherently with the incident laser beam. Schematic of the `wind-chime' model has been presented in Figure {\ref{fig 3:3}(a)}.  By this process, the whole system produces the SSPM through the nonlocal coherent interaction between the MoSe$_2$ NFs and incident laser beams.

Figure {\ref{fig 3:3}(b-d)} shows the evolution process of the SSPM of the incident light beams with different wavelengths propagating through the 2D MoSe$_2$ NFs. Initially, a bright spot has formed in the far-field. Gradually, a fully circular diffraction pattern emerges as the NFs become aligned along the optical field. After some time, the upper half of the diffraction rings start to collapse and slowly becomes stable. The evolution of diffraction rings over time is presented in Figure {\ref{fig 3:3}(b-d)} for laser beams with $\lambda$ = 671, 532 and 405 nm, respectively. According to the wind chime model, the time required to form the maximum diameter full circular diffraction pattern is equal to the time required to reorient these NFs along the electric field of the incident polarized light. The time evolution of the diffraction ring numbers is presented in Figure \ref{fig 3:3}(e), and the dynamics of formation diffraction rings follow the exponential model as,

\begin{equation}
	N= A~(1~-~e^{-t/\tau_c})
	\label{eq:decay}
\end{equation}

where $N$ is the number of rings, $\tau_c$ is the rise time for the ring formation, and $A$ is a constant. As shown in Figure {\ref{fig 3:3}(e)}, the  $\tau_c$ is estimated to be 0.209, 0.317 and 0.0393 s for 671, 532 and 405 nm, respectively. The minimum time ($\mathcal{T}$) required for reaching the highest number of rings is estimated to be 0.41, 0.62 and 0.74 s, respectively. According to the wind-chime model, the time required for the pattern formation is as follows \cite{Wu2015, Zhang2016},
\begin{equation}
	\begin{aligned}
		\mathcal{T} = \frac{\epsilon_r\pi\eta\xi Rc}{1.72(\epsilon_r -1)Ih} 
	\end{aligned}
	\label{eq:evolution}
\end{equation}

where $\epsilon_r$ is the relative dielectric constant of MoSe$_2$, $\eta$ is the coefficient of viscosity of the solvent, $R$ is the domain radius ($\sim$ 100 nm), $h$ is the flake thickness ($\sim$ 5 nm), and $I$ is the laser intensity ($\sim$ 12 W/cm$^2$).  As we know that the formation time of diffraction ring (equation no\ref{eq:evolution}) strongly depends on the size, thickness, and dielectric constant of the materials. Also, $\mathcal{T}$ depends upon the viscosity ($\eta$) of the dispersion medium, and they are proportional to each other.  Hence, keeping all the parameters same, if we use solvents with different viscosity, the formation time of those diffraction rings are also different. Low viscous medium requires a shorter time to reorient the 2D NFs towards incident electric field rather than high viscous medium. To verify the wind-chime model, two different viscous medium like NMP ($\eta$ = 1.65 $\times$ 10$^{-3}$ Pa.s) and acetone ($\eta$ = 3.2 $\times$ 10$^{-4}$ Pa.s) have been chosen. As expected, the theoretically estimated $\mathcal{T}$ is found to be 0.38 and 0.26 s for NMP and acetone solvent, respectively. Also, the $\mathcal{T}$ is estimated from the experimental data (\ref{fig 3:3}(f)) to be 0.41 and 0.22 s for NMP and acetone solvent, respectively. The experimental results are in good agreement with the previously reported values for other materials, as presented in table \ref{tab:2}. \\
\begin{table}
	\centering
	\caption{Diffraction ring formation time from `wind-chime' model}
	\label{tab:2}
	\begin{tabular}{ccccccc}
		\hline
		&Materials &Solvent  &Wavelength &Intensity &Formation time &Reference \\
		& & &$\lambda$ (nm) &$I$ (cm$^2$/W) &$\mathcal{T}$(s)& \\
		\hline
		&MoS$_2$ &NMP & 532 & 250 &0.2 &\cite{Wu2015} \\
		&MoTe$_2$ &NMP & 750 & 252 &0.62 &\cite{Hu2019} \\
		&Black phosphorus &NMP & 700 & 18.9 &0.7 &\cite{Zhang2016} \\
		&Graphite &NMP & ---  & 100 &0.43 &\cite{Wu2016} \\
		&MoSe$_2$ &NMP & 671 & 12 &0.41 &This \\
		& &Acetone & 671 & 12 &0.22 &work \\
		\hline
	\end{tabular}
\end{table}%

\medskip

To investigate the effect of polarity of the solvent, we have used one polar (NMP) and one non-polar (toluene) solvents keeping other parameters constant. The results are presented in Figure \ref{fig 3:3}(g). In both cases,  similar diffraction patterns are observed, which confirms that these NLO effects are truly the properties of 2D semiconducting materials, and the nature of polarity of the solvent doesn't play any role. \\

In addition to the Kerr nonlinearity, the change in the temperature of the medium under an intense laser beam can also modify the refractive index of the medium and produce similar self-phase modulation as described above. The above phenomenon is known as the `thermal lens effect' \cite{Dabby1970}. The SSPM is an NLO phenomenon and depends on the reorientation or polarization of the materials with an incident laser beam. In contrast, the thermal lens effect has a linear optical response \cite{Dabby1970, Wang2017}. To confirm the origin of the diffraction rings, a mechanical chopper was introduced into the path of the incident laser beam \cite{Liao2019}. In our experiments, the chopper can modulate the intensity of the laser at a frequency of 20~Hz to 400 Hz \cite{Wu2020a, Shan2019}. Figure \ref{fig 3:3}(h) (\textcircled{1}-\textcircled{6}) shows the diffraction ring patterns for modulation frequency, $f$ = 0, 20, 50, 100, 200 and 400 Hz, respectively. The above results show that the number of diffraction rings after using the chopper is less than those without a chopper. These observations confirm that the NLO response is the inherent feature and dominates the thermal lens effect. 

\begin{figure}[t]
	\centering
	\includegraphics[width=0.75\columnwidth]{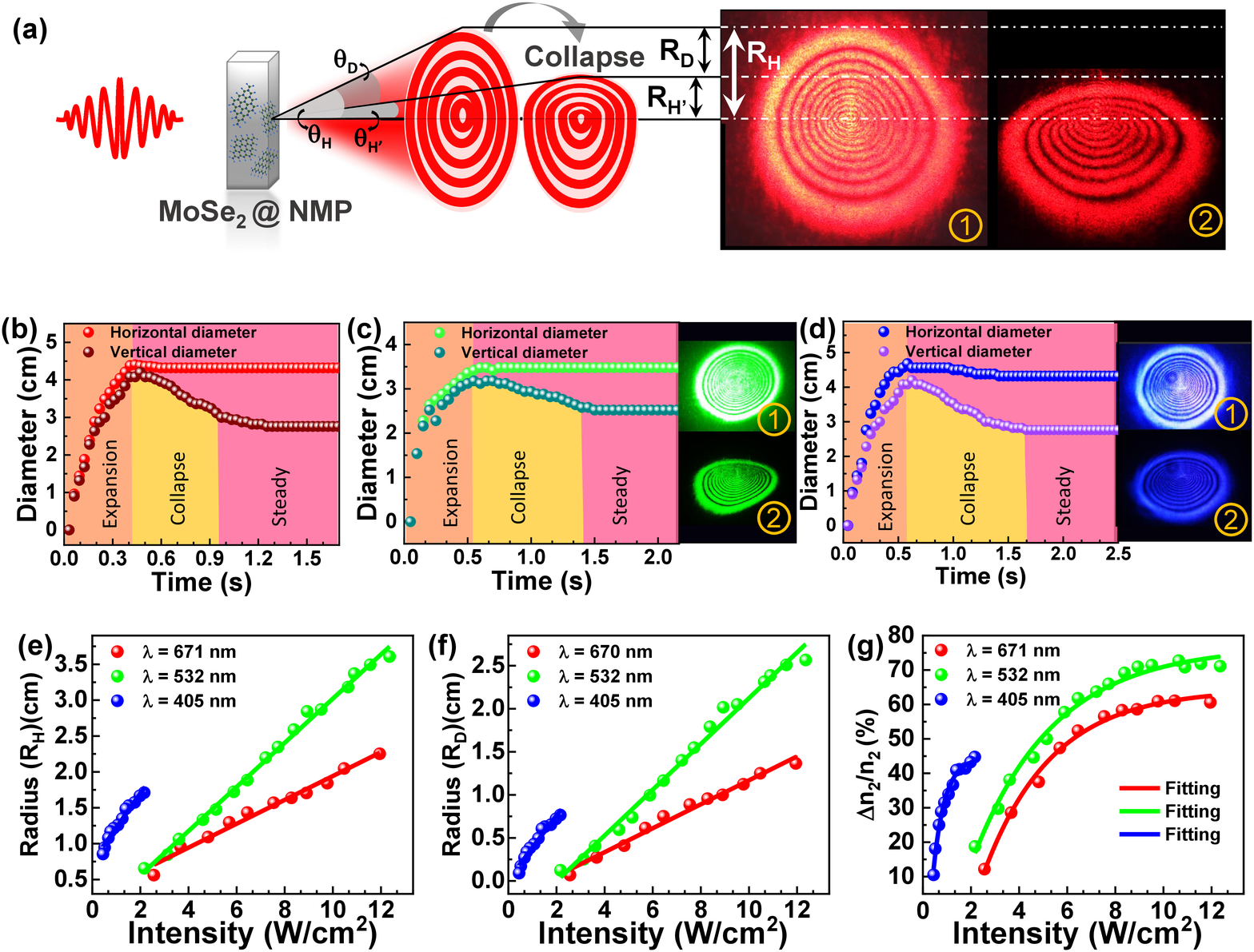}
		\caption{\textbf{Collapse phenomena of SSPM diffraction patterns.} (a) Schematic presentation of the collapse process of the diffraction ring  with half-cone and distortion angle. \textcircled{1} and \textcircled{2} are the real images of the SSPM diffraction pattern at maximum and distorted condition. (b-d) The evolution of the vertical and horizontal diffraction ring diameters over time for incident laser light of $\lambda$ = 671, 532, and 405 nm, respectively. The variation in the (e) maximum radius (R$_H$) and (f) distorted radius (R$_D$) of the diffraction ring for different incident laser of $\lambda$ 671, 532 and 405 nm.  (g) The relative change in the nonlinear refractive index with incident light intensity for $\lambda$ = 671, 532, and 405 nm.}
	\label{fig 4:4}
\end{figure}

\subsection{Dynamics of collapse phenomena of diffraction patterns}

After a close look, it has been observed that the diffraction rings are getting distorted with time. In the SSPM experiment, the diffraction ring gradually develops once the focused laser passes through MoSe$_2$ NFs. After the diffraction ring reaches its maxima, the upper half of the vertical radius begins to collapse towards the center. More distortion occurs along the vertical direction compared to the horizontal direction, as shown in Figure \ref{fig 4:4}. To be more precise, the lower half of the diffraction rings remained stable, whereas the upper half collapses. After a few seconds, it reaches the equilibrium situation as shown in Figure \ref{fig 4:4} (a). This collapse (upper half) process is due to the non-axis-symmetrical thermal convection leading to the distortion of diffraction pattern described by Wang et al. \cite{Wang2014}. In this process, the dispersive medium absorbs some part of the propagating laser light due to its finite optical absorption coefficient\cite{Wang2009, Wu2011}. As a result, the temperature gradient increases along the vertical direction around the laser spot, enhancing the thermal convection process\cite{Teng2019}. The above convection process of the medium reduces the concentration of the 2D materials in the upper part of the medium than the lower part. Therefore, the upper part of the laser beam is less diffracted by the dispersed 2D materials leading to distorted diffraction patterns which collapse towards the center. The above phenomenon is quantified by the half-cone angle of the diffraction pattern of the Gaussian laser beam, which can be expressed as \cite{Wang2015}}

\begin{equation}
	\theta_H = n_2  \left[ -\frac{8IrL}{w^2_0} exp\left( -\frac{2r^2}{w^2_0}\right) \right]  _{max}
	\label{eq:theta_H}
\end{equation}

where $\left[ -\frac{8IrL}{w^2_0} exp\left( -\frac{2r^2}{w^2_0}\right) \right]$ is a constant when $r \in [0,+\infty)$. As a result, $\theta_H$ is proportional to the effective nonlinear refractive index of the dispersed 2D materials. 

Figure \ref{fig 4:4}(a) shows the schematic presentation of the time-dependent collapse process by changing the diffraction angle. The intensity-dependent collapse behaviour of the diffraction rings can be quantified by the change of diffraction angle formed between the sample and the screen \cite{Li2020}. These diffraction rings are formed through a series of coaxial cones, and the distortion angles are presented as \cite{Jia2018, Wu2018, Zidan2020}\\

\begin{equation}
	\begin{aligned}
		\theta_D = \theta_H - \theta_{H'} = \frac{R_H}{D} - \frac{R_{H'}}{D} =\frac{R_D}{D}
	\end{aligned}
	\label{eq:distortion}
\end{equation}

where $R_H$ is the maximum diffraction radius, $\theta_H$ is the maximum half diffraction angle. After the distortion, the two parameters $R_H$ and $\theta_H$ must be changed to new values of $R_H'$ and $\theta_H'$, respectively, and $\theta_D$ can also be described as a difference of nonlinear refractive index as  

\begin{equation}
	\begin{aligned}
		\theta_D = \theta_H - \theta_{H'} = (n_2-n_2')IC = \Delta n_2IC
	\end{aligned}
	\label{eq:del n}
\end{equation}

Therefore, the final relation between the relative change in nonlinear refractive index can be presented in terms of the ratio between the distorted and maximum half-angle as \cite{Liao2020a}

\begin{equation}
	\begin{aligned}
		\frac{\Delta n_2}{n_2} = \frac{\theta_D}{\theta_H} = \frac{R_D}{R_H}
	\end{aligned}
	\label{eq:del n vs theta}
\end{equation}

Based on the above equation \ref{eq:del n vs theta}, the distortion process of the diffraction rings is qualitatively analyzed by $\Delta$n$_2$/n$_2$ with the dynamic change of the distortion and full radius of the diffraction rings ($R_D/R_H$). Typically, the $\Delta$n$_2$/n$_2$ is mainly depends on the laser intensity, temperature and time \cite{Liao2020a}. Among the above three parameters (intensity, temperature, and time), the intensity of the laser beam has the most prominent dependence on the change of the nonlinear refractive index (n$_2$) \cite{Wang2015, Jia2019}.

Figure \ref{fig 4:4}(a)(\textcircled{1}-\textcircled{2}) shows the image of the diffraction ring patterns after attaining the maximum size and at the steady-state condition after the collapse process for the laser beam with $\lambda$ = 671 nm. Figure \ref{fig 4:4}(b-d) shows the evolution of the diffraction ring pattern over time for all three laser beams with $\lambda$ = 671, 532, and 405 nm, respectively, at their highest intense laser power. For all three lasers, the horizontal diameters increased to the maximum diameter during the expansion and remained constant for the rest of the time. At the same time, the vertical diameters reach the maximum size and then slowly collapse with time to steady-state values. Interestingly, the collapse time increases (0.54, 0.89, and 1.06 sec) with the increasing photon energy of the laser beam ($\lambda$= 671, 532, and 405 nm). The experimentally measured maximum vertical radius (R$_H$) before the collapse and the radius after the collapse process (R$_D$) are presented in Figure \ref{fig 4:4}(e-f). Here, during the collapse process, the deformation of the diffraction rings are observed due to non-axis-symmetrical thermal convection as described above.\\

Figure \ref{fig 4:4} (g) shows the dependence of relative change in the nonlinear refractive index with the incident laser beam intensity using the eq \ref{eq:del n vs theta}.  The $\Delta n_2 / n_2$ increases with increasing laser light intensity, but the relative change tends to be saturated with the increasing laser light intensity. The nonlinear effect of blue light is the strongest. The relative nonlinear refractive index $\Delta n_2 / n_2$ of the MoSe$_2$ NFs can be found to 40\% (2W/cm$^2$), 75\% (12.5W/cm$^2$)and 60\% (12W/cm$^2$) for the incident light with $\lambda$= 405, 532 and 671 nm respectively.
 
\subsection{Nonreciprocal light propagation in MoSe$_2$ based photonic diode}

\begin{figure} [t]
	\centering
	\includegraphics[width=0.65\columnwidth]{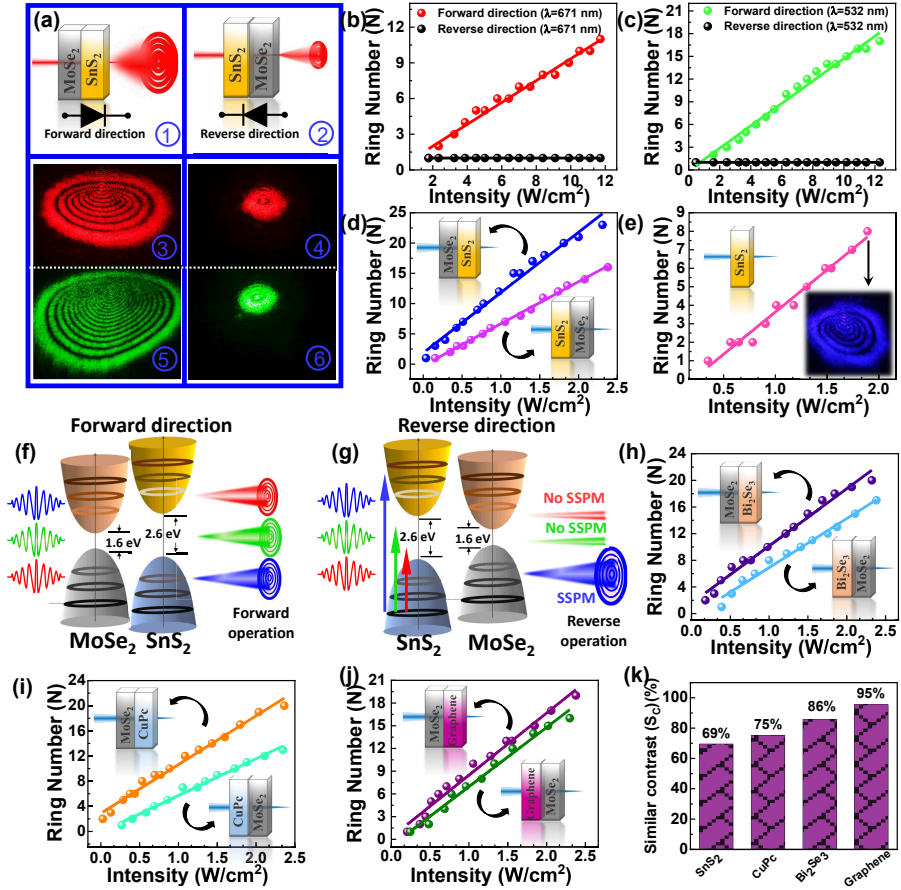}
	\caption{\textbf{Nonreciprocal light propagation in all-optical photonic diode.} (a)~\textcircled{1}~-~\textcircled{2} shows the schematic presentation of the nonreciprocal light propagation in forward and reversed direction. \textcircled{3} and \textcircled{5} shows the diffraction patterns obtained at forward condition for $\lambda$ = 671, and 532 nm, respectively.  \textcircled{4} and \textcircled{6} shows the diffraction rings obtained at reversed condition for $\lambda$ = 671, and 532 nm, respectively. The experimental result obtained for the nonlinear photonic diode based on MoSe$_2$/SnS$_2$ hybrid structure for incident laser light of (b) $\lambda$ = 671 nm and (c) $\lambda$ = 532 nm. (d) The diffraction ring numbers as a function of intensity of the incident light $\lambda$ = 405 nm for the MoSe$_2$/SnS$_2$ hybrid structure. (e) The SSPM effect for independent SnS$_2$ dispersed in NMP excited by $\lambda$ = 405 nm. (inset shows the diffraction rings at highest intensity). (f) and (g) Schematic diagram of the microscopic band diagram for all-optical forward and reverse photonic diode operation, respectively. The diffraction ring number variation for both forward and reverse direction with incident light intensity for the hybrid structure of (h) MoSe$_2$/Bi$_2$Se$_3$ , (i) MoSe$_2$/CuPc (j) MoSe$_2$/graphene, respectively. (k) The similar contrast (S$_C$) of MoSe$_2$ comparing to other semiconducting materials (SnS$_2$, CuPc, Bi$_2$Se$_3$, graphene and Sb$_2$S$_3$).}
	\label{fig 5:5}
\end{figure}

Here we demonstrate a novel nonlinear photonic diode device using MoSe$_2$/SnS$_2$ hybrid structure based on the SSPM. The SnS$_2$ is another nonlinear material that shows strong NLO response and SSPM. The SnS$_2$ has a bandgap (2.6 eV) larger than MoSe$_2$ (1.6 eV) and the hybrid structure can be used as a photonic diode for nonreciprocal light propagation\cite{Song2020, Wu2019}. The laser lights with $\lambda$ = 532 and 671 nm are used to characterize the photonic diode based on MoSe$_2$/SnS$_2$ hybrid structure. The photonic diode produces the diffraction pattern when the laser beams propagate in the forward direction (i.e. MoSe$_2$$\rightarrow$SnS$_2$) as shown in Figure \ref{fig 5:5} (a) \textcircled{2},\textcircled{4},\textcircled {6}. Whereas in the reverse direction (i.e. MoS$_2$$\leftarrow$SnS$_2$), no such diffraction rings are observed as shown in Figure \ref{fig 5:5} (a) \textcircled{1},\textcircled{3},\textcircled {5}. In the reverse direction large absorption through SnS$_2$ reduces the intensity of the laser in MoS$_2$. Therefore, the effect of SSPM is negligible and there is no diffraction patterns. Therefore the above properties of  proposed hybrid photonic diode can be used for all-optical switch applications where the nonreciprocal propagation of light can be achieved. 

Laser light intensity-dependent diffraction ring numbers with unidirectional property of  MoSe$_2$/SnS$_2$ photonic diode for laser beams with $\lambda$ = 671 and 532 nm, are presented in Figure \ref{fig 5:5} (b) and (c). Here the values of dN/dI are found to be 1.479 and 0.9186 for $\lambda$ = 532 and 671 nm, respectively. The values are very much comparable with single MoSe$_2$ results. The photonic diode has a higher NLO response at $\lambda$=532 nm than $\lambda$=671 nm, and it can perform better as an optical diode at a higher wavelength.

To further check the unidirectional performance of the nonlinear photonic diode (using $\lambda$ = 405 nm) in forward and reverse operation and results are presented in Figure \ref{fig 5:5} (d).  Interestingly, the diffraction rings appear in both directions, but the number of diffraction rings is different. Therefore, the MoSe$_2$/SnS$_2$ photonic diode is not appropriate for lower $\lambda$ like 405 nm, and to investigate it, we have performed the SSPM for only SnS$_2$ using 405 nm laser and results are shown in Figure \ref{fig 5:5}(e). As expected, the SSPM is observed in SnS$_2$ (bandgap $\sim$ 2.6 eV) when the laser with higher excitation energy (2.71 eV) is applied \cite{Liao2019}.  To understand the basic mechanism for photonic diode application through the light-matter interaction, schematically, we present the energy band for both MoSe$_2$ and SnS$_2$ in Figure \ref{fig 5:5} (f \& g) for forward and reverse, respectively. 

 In this experiment, three different lasers having higher energy than MoSe$_2$ (bandgap $\sim$ 1.6 eV) are used, and a photon of energy $E = \hbar\omega$ can excite the electrons from the valance band to the conduction band. The excited electrons go to the ground state by releasing the photon. The emitted photons interact with the incident light and produce the diffraction rings through the optical Kerr effect \cite{Wu2019, Liao2019}. In this process, the excited electrons will move antiparallel with the applied optical electric field (coming from the laser beam) and polarize the suspended MoSe$_2$ NFs. Next, the reorientation process reduces the angle between the polarization direction and the external electric field to achieve the minimum interaction energy configuration. That increases the NLO response of 2D MoSe$_2$ NFs, leading to the nonlinear Kerr effect \cite{Li2020, Wang2015}. 

 However, 532 and 671 nm lasers are unable to excite the electrons in SnS$_2$ from valance band to conduction band and mostly get absorbed due to the intraband transitions. Therefore, no such diffraction rings are observed \cite{Wu2019, Wu2020, Wu2020a}. The laser beam with $\lambda$ = 405 nm produces diffraction rings in both the pure SnS$_2$ system as well as the MoSe$_2$/SnS$_2$ photonic diode.  

To further estimate the $n_2$ of the photonic diode MoSe$_2$ and few other semiconducting materials from the nonreciprocal light propagation characteristics, similar experiments were preformed. The similarity comparison method (SCM) allow us to estimate similarity contrast ($S_C$) of these materials using their nonlinear refractive indices as \cite{Wu2019},

\begin{equation}
	\begin{aligned}
		S_C= 1-D_C  
		&=1- \frac{\big|n_{21}-n_{12}\big|}{n_{21}} \\
		&=1-\frac{\big|\frac{\lambda}{2n_0L_{eff}}\frac{dN_1}{dI_1}-\frac{\lambda}{2n_0L_{eff}}\frac{dN_2}{dI_2}\big|}{\frac{\lambda}{2n_0L_{eff}}\frac{dN_1}{dI_1}}\\
		&= 1- \frac{\big|\frac{N_1}{I_1}-\frac{N_2}{I_2}\big|}{\frac{N_1}{I_1}}
	\end{aligned}
\end{equation}  
where $D_C$ is the difference constant, $n_{21}$ and $n_{22}$ represent the nonlinear refractive index of the hybrid system obtained for the forward and reverse direction respectively. The semiconducting  Bi$_2$Se$_3$, CuPc, and graphene are used in combination with MoSe$_2$ for $S_C$ studies which has been presented in Figure \ref{fig 5:5}(h-j) and summary of the results is presented in Figure \ref{fig 5:5}(k). The $S_C$ values for SnS$_2$, CuPc, Bi$_2$Se$_3$ and graphene  are found to be 69\%, 75\%, 86\% and 95\%,  respectively. This results indicate that the MoSe$_2$ has similar nonlinear refractive index to the graphene \cite{Wu2019, Wu2016, Shan2019a, Wang2014}. Furthermore, the CuPc has very low $S_C$ which is close to SnS$_2$ for $\lambda$ = 405 nm. \\

\subsection{Cross-phase modulation: MoSe$_2$~- based all-optical switching and logic gates }
\begin{figure} [t]
	\centering
	\includegraphics[width=0.65\columnwidth]{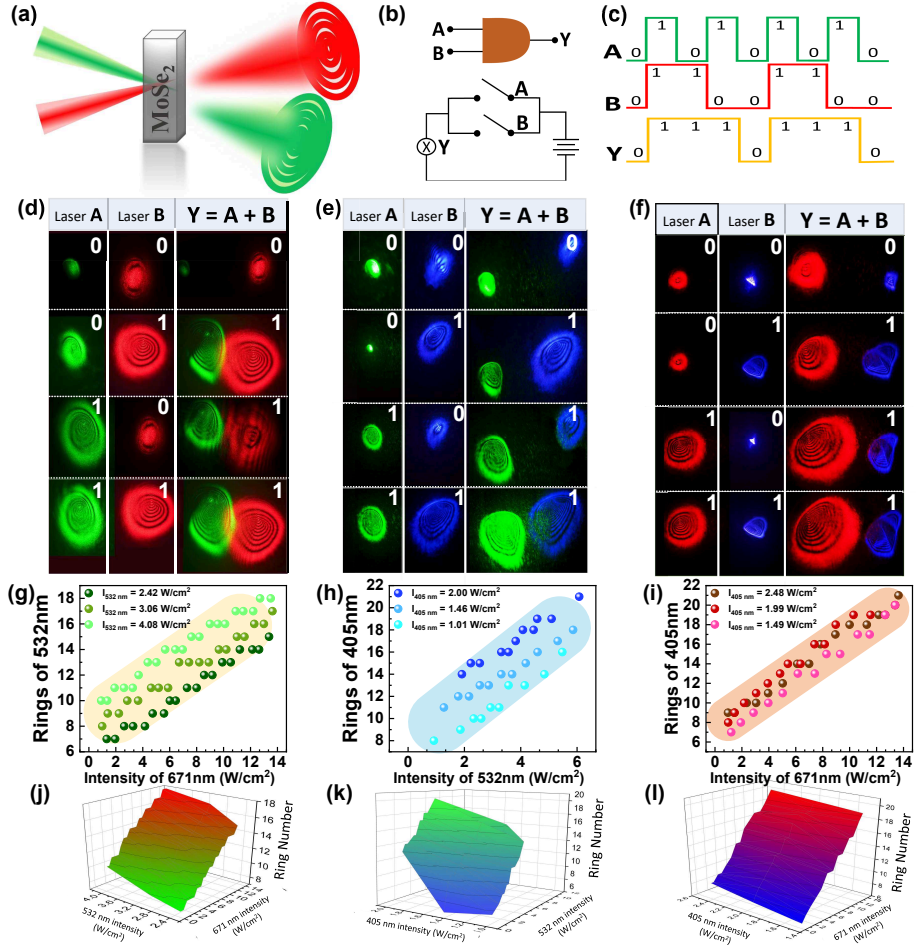}
	\caption{\textbf{Cross-phase modulation: All-optical switching and logic gates.} (a) Schematic presentation of the XPM inside MoSe$_2$ NFs for photonic diode application. (b) Symbol and circuit model of the logical OR gate. (c) The waveform of OR gate (input: $A$ and $B$; output: $Y$) (d-f) MoSe$_2$ based experimental results for optical OR gate operation using two-color XPM technique for the combination of $\lambda$= 532 \& 671 nm,  405 \& 532 nm, and 405 \& 671 nm, respectively.  (g-i) Change in the diffraction ring numbers of the probe lasers vs the intensity variation of the pump laser. (j-l) Three dimensional presentation of the two-color XPM for different pair of lasers ($\lambda$= 532 \& 671 nm,  405 \& 532 nm, and 405 \& 671 nm).}
	\label{fig 6:6}
\end{figure}

By taking advantage of the SSPM by MoSe$_2$ NFs, an all-optical switch has been demonstrated to perform the logical function like an OR gate as shown in Figure \ref{fig 6:6}. The schematic of the experimental logical OR gate is presented in Figure \ref{fig 6:6} (a) using the cross-phase modulation (XPM) technique \cite{Lu2017, Shan2019b}. Two lasers with different wavelengths are used as the two input signals ("A" and "B") of the logic system on the MoSe$_2$ NFs based logic device. The resultant XPM are the corresponding outputs of the logic gate. Figure \ref{fig 6:6}(b) shows the symbol and switch circuit of the OR logic gate. Where the input "A" or "B" is in high-level "1", the output "Y" will also be in high-level "1", and only when both are at the lower level "0", the output will be low "0". The corresponding switching waveforms are ("A", "B", and "Y") presented in Figure \ref{fig 6:6} (c). Here, all the logic levels are realized by the number of diffraction rings formed using the XPM process. Here, we observed a two-laser-based system to achieve light-light modulation. The diffraction rings cannot be formed when the incident light or probe light is too low to interact with the sample, and in this case, only a Gaussian light appears on the screen. When another laser light (pump light) with relatively high intensity is cross-modulated with the Gaussian probe light, it can form diffraction rings on the screen for both the lights. The individual lights of $\lambda$= 532 nm and 671 nm are considered input "A" and input "B", respectively. Now both lights are cross-coupled, and the final image is formed on the screen, which is considered as the output "Y". Based on the output using two-lasers, we can build an optical logic OR gate, which is presented in Figure \ref{fig 6:6} (d-f) for the combination of laser lights with $\lambda$= 532 \& 671 nm,  405 \& 532 nm, and 405 \& 671 nm, respectively. Figure \ref{fig 6:6}(g)  presents the number of diffraction rings of the probe light ($\lambda$=532 nm) with the intensity signal of the pump light $\lambda$ = 671 nm, where the probe light intensities are 2.42, 3.06 and 4.08 $W/cm^2$, respectively. The initial number of diffraction rings formed by the probe light depends on the intensity of the probe light. The diffraction rings are modulated based on the superposition principle when the pump light is on. In addition, the diffraction ring numbers of the probe light linearly depend on the pump light's intensity. For other two set of light combinations,  $\lambda$ =405 \& 532 nm, and 405 \& 671 nm,  the similar experiments are performed and presented in Figure \ref{fig 6:6} (g) and (i), respectively. Figure \ref{fig 6:6} (j-l) shows the three dimensional model of the overall data obtained through the diffraction ring numbers of the probe light with the intensity of the pump and probe light for choice of pump and probe lights with $\lambda$= 532 \& 671 nm,  405 \& 532 nm, and 405 \& 671 nm, respectively. \\

\section{Conclusions}
In conclusion, we have reported the spatial self-phase modulation (SSPM) of Gaussian laser beams by suspended MoSe$_2$ nanoflakes (NFs) in a solvent. The NFs of 2D layered MoSe$_2$ are synthesized using a cost-effective solvothermal technique. The nonlinear broadband optical response of MoSe$_2$ NFs has been investigated using laser beams with $\lambda$ = 671, 532, and 405 nm. The nonlinear refractive index and the monolayer third-order nonlinear susceptibility are estimated from the change in the diffraction ring number as a function of laser power. The obtained n$_2$ and $\chi^{(3)}_{(monolayer)}$ for MoSe$_2$ NFs are found to be $1.3 \times 10^{-5}$ cm$^2$/W and $2.23 \times 10^{-8}$ esu, respectively at 532 nm excitation laser, are comparatively larger than other family members of TMDs. The large values of n$_2$ and $\chi^{(3)}$ reveal strong nonlinear coherent interaction of light with the ensemble of layered 2D MoSe$_2$ NFs. The evolution (formation) of the self diffraction ring patterns due to SSPM have been described by the `wind-chime' model considering coherent light-matter interactions. Based on the SSPM, a passive photonic diode has been designed using MoSe$_2$/SnS$_2$ hybrid heterostructures to demonstrate the nonreciprocal light propagation. The unidirectional properties of the above photonic diode using MoSe$_2$ have been compared with a few other semiconducting materials (i.e., Bi$_2$Se$_3$, CuPc, and graphene) by the similarity comparison method. The all-optical switching properties have also been explored based on the two-color interband coherence. In addition, the all-optical information processing has been examined by performing the "OR" logic gate operation using cross-phase modulation (XPM) of two laser beams. We believe that these NLO phenomena can provide an inroad foundation for the MoSe$_2$ based all-optical switching as optical diodes and logic devices.

\section{Experimental Section}
\label{sec:Experimental}
\threesubsection{Synthesis of MoSe$_2$ nanoflakes}\\
Facile solvothermal technique has been adapted to synthesize MoSe$_2$ nanoflakes (NFs), and Figure \ref{fig 1:1} (a) shows the schematic presentation of the synthesis process. The detailed preparation method of MoSe$_2$ NFs has been discussed in our earlier works \cite{Maity2020, Das2021}. In brief, a suitable amount of sodium molybdate ($Na_2MoO_4.2H_2O$) was added in 22.5 ml dimethylformamide (DMF) solution with 30 min sonication for proper dispersion at room temperature. Next,15 mg of selenium powder is dispersed in 7.5 ml hydrazine hydrate by constant stirring in an oil bath at 80$^\circ$C for 1h. Then, this solution was mixed with the previously prepared DMF solution to maintain the atomic ratio of Mo: Se is 1:2. Then, the as-prepared solution was transferred into a 100 ml autoclave up to 60 \% of its volume and kept into an oven, which was already preheated at 180$^\circ$C  for the next 12 h. Thereafter, the autoclave was cooled down by natural cooling. Next, the obtained precipitate was filtered and washed by de-ionized water and ethanol multiple times. Finally, the black MoSe$_2$ powder was annealed at 450$^\circ$C under $N_2$ atmosphere for 4 h to improve the crystallinity of MoSe$_2$ NFs. \\

\threesubsection{Synthesis of SnS$_2$ nanosheets}\\
The synthesis process of SnS$_2$ has been done following process reported by Zhang \textit{et al.}\cite{Zhang2019a}. Typically, 220 mg $SnCl_4.5H_2O$ and 280 mg thioacetamide were mixed with 60 mL of de-ionized water and
then transferred to a 100 mL Teflon-lined autoclave and the autoclave was heated and kept at 180$^{\circ}$C for 24 hours. After the reaction, the autoclave was allowed to cool naturally to room temperature, and the yellow-colored sample was collected after centrifuging the sample at 10000 rpm. The final product was dried in a hot air oven at 80$^\circ$C for 6 hours and stored in a desiccator. The detailed characterization has been carried out and discussed in the section S1 of supplementary.

\threesubsection{Characterization}\\
Field-emission scanning electron microscope (FESEM Hitachi S-4800) and transmission electron microscopy (FEG-TEM) were used to characterize the sample morphology. Figure \ref{fig 1:1} (b) and (c)  shows the layered MoSe$_2$ NFs.  TEM-EDX (transmission electron microscopy-energy-dispersive X-ray spectroscopy) spectrum and elemental mapping reveals the chemical composition and stoichiometric ratios of MoSe$_2$ NFs (see Figure\ref{fig 1:1}(d)). The above analysis confirms that the sample contains the elements Mo and Se with their atomic ratio is about 1$:$~2.07. Figure \ref{fig 1:1} (e) shows the Raman spectrum of 2D MoSe$_2$ NFs with its characteristics in-plane (E$_{1g}$, E$^1_{2g}$) and out-of-plane (A$_{1g}$) vibrational modes. The high-resolution X-ray photoelectron spectroscopy (XPS) shows the chemical environment of an element Mo 3d and Se 3d, which are presented in Figure \ref{fig 1:1} (f) and (g), respectively. The bandgap of the MoSe$_2$ NFs sample was determined to be Eg = 1.6 eV by the UV-Vis absorption curve and corresponding Tauc plot curve as shown in Figure \ref{fig 1:1} (h). The characterized 2D layered MoSe$_2$ NFs have been dispersed in NMP (N-methyl- 2-pyrrolidone) and a few other solvents (acetone, toluene, etc.) with different loading concentrations for the SSPM experiments. 

\threesubsection{Experimental setup}\\
 The schematic of SSPM spectroscopy setup to study the NLO response of MoSe$_2$ NFs suspended in NMP solvent is shown in Figure {\ref{fig 2:2}}(a). The sample shows the broadband SSPM effect for the incident pump light with $\lambda$ = 671, 532, and 405 nm, respectively. Here, the pump laser passes through the convex lens of 20 cm focal length ($f$) and incident on a quartz cuvette of the thickness of 10 mm,  which is placed near the focal point of the lens. Under intense coherent Gaussian laser beam, the suspended MoSe$_2$ NFs interact with the incident  laser and the MoSe$_2$ NFs start to orient along the optical electric field \cite{Xiao2020a}.

\medskip
\textbf{Supporting Information} \par
Supporting Information is available from the Wiley Online Library or from the author.

\medskip
\textbf{Acknowledgements} \par
This work is supported by the `Department of Science and Technology' under startup research grant (Grant No. SRG/2019/000674). One of the authors, K. K. Chattopadhyay acknowledges the University Grants Commission, Govt. of India, under the 'University with potential for excellence II' (UPE II) scheme. The author, MS gratefully acknowledges DST 'INSPIRE' scheme [IF170868] for giving research opportunities through fellowship. AB \& SB thanks CSIR Govt. of India for Research Fellowship with Grant No. 09/080(1109)/2019-EMR-I \& 09/080(1110)/2019-EMR-I, respectively. The authors want to thank Dr Sudipta Bera for his valuable support in preliminary discussions and setting-up the experiment. His contribution is very much appreciated for this project.

\medskip
\textbf{Conflict of Interest} \par
The authors declare no competing financial interest.


\bibliographystyle{unsrt}
\bibliography{SSPMRefDB}
\bigskip
\newpage
\hrule

\hfill

\bigskip

\textbf{Supporting Information: Nonlinear coherent light-matter interaction in 2D MoSe$_2$ nanoflakes for all-optical switching and logic applications} \par
\bigskip

\textbf{$\bullet$ Laser induced SSPM in optically nonlinear materials}  
\bigskip\\
2D materials are drawing much more attention in the field of nonlinear optical (NLO) which can be quantitatively described using basic Electrodynamics and theories of light-matter interactions. Generally, the NLO effect occurs when an intense laser beam modulates the optical properties of the material. The optical responses of the materials are reacted in a nonlinear manner with the power of the incident laser beam. To visualized this effect mathematically, the applied electric field ($\tilde{E}(t)$) can be expressed by expanding the resulting polarization ($\tilde{P}(t)$), has both the linear and the nonlinear contributions, given by equation\ref{eq:polarization}:
\begin{equation}
	\begin{split}
		\tilde{P}(t) 
		& = \tilde{P_{L}}(t) +  \tilde{P_{NL}}(t) \\
		&= \tilde{P}_L^{(1)}(t)+\tilde{P}^{(2)}(t)+\tilde{P}^{(3)}(t)+.....\\
		&=\epsilon_0\big[\chi^{(1)}\tilde{E}(t)+\chi^{(2)}\tilde{E}^2(t)+\chi^{(3)}\tilde{E}^3(t) + ....\big]\\
	\end{split}
	\label{eq:polarization}
\end{equation}

where  $\epsilon_0$ is the vacuum permittivity and the coefficients $\chi^{(n)}$ are the $n^{th}$- order susceptibilities of the materials. The first coefficient $\chi^{(1)}$  is the linear susceptibility and also it is a tensor of rank two. $\chi^{(1)}$ describes the linear optical responses, such as absorption and refraction of the materials. The quantities $\chi^{(2)}$ and $\chi^{(3)}$ are known as the second-and third-order NLO susceptibilities, respectively, which are normally much weaker than $\chi^{(1)}$. When the optical field is strong enough, the phenomena caused by higher-order (n$ \geqslant$2) terms become significant, which could induce new radiation at different frequencies, pulse shaping, or change of refractive index \cite{Boyd2020}.
The emergence of nonlinear optics has been the focus of modern research after the breakthrough discovery of the second-harmonic-generation(SHG) in the strong-optical-field regime. Numerous phenomena in condensed matter systems is probed in past decades using SHG technique \cite{Clark2015, Xia2021}. \\
The third-order NLO susceptibility $\chi^{(3)}$ can also give rise different NLO effects, such as third-harmonic-generation (THG), four-wave mixing (FWM), optical Kerr effect (leading to self-phase modulation), and saturable absorption (SA). Although higher-order NLO interactions are much weaker than second- and third- order NLO, and appears at extremely high laser intensity \cite{Shen1984}.

\bigskip
\textbf{Theory of self-phase modulation and spectral broadening} \par
\bigskip

NLO phenomena can be quantitatively described using Electrodynamics and theories of light-matter interaction. For self-phase modulation and spectral broadening when the propagation of a laser beam in an isotropic medium can be described by the wave equation of a plane wave:

\begin{equation}
	\begin{split}
		\left[ \nabla^{2}  - \frac{n_0^2}{c^2}\frac{\partial^{2} }{\partial t^2}\right]  \tilde{E}(t) &= \frac{1}{\epsilon_0 c^2}\frac{\partial^{2} \tilde{P}_{NL}(t)}{\partial t^2}\\
	where, ~~~	\tilde{E}(t) &= \tilde{E}_0(\omega)\big[e^{+i\omega t} +e^{-i\omega t}\big]\\
and,  ~~~	\tilde{P}^{(3)}_{NL} &= \epsilon_0 \chi^{(3)} \tilde{E}^3_0(\omega)\big[e^{+i\omega t} +e^{-i\omega t}\big]^3
	\end{split}
		\label{eq:Elcetrosynamics}
\end{equation}

where$ \tilde{E}(t)$ is the optical field, $n_0$ is the linear refractive index, and $c$ is the speed of light in a vacuum.  $\tilde{P}^(3)_{NL}$ of a materials gives rise to optical phenomena like THG and intensity dependent refractive index. Time domain nonlinear polarization function  ($\tilde{P}^3_{NL}(t)$) is shown as Eqn(\ref{eq:response})

\begin{equation}
	\tilde{P}^(3)_{NL}(r,t) = \epsilon_0\int_{-\infty}^{+\infty} \int_{-\infty}^{+\infty}\int_{-\infty}^{+\infty}R^3((z-t_1),(t-t_2),(t-t_3))\tilde{E}(\vec{r},t_1)\tilde{E}(\vec{r},t_2)\tilde{E}(\vec{r},t_3) \,dt_1\,dt_2\,dt_3
	\label{eq:response}
\end{equation}

Therefore, the effective susceptibility  ($\chi_{eff}$) can be derive from the equation \ref{eq:polarization} and  \ref{eq:Elcetrosynamics} as ,
\begin{equation}
	\chi_{eff} = \chi^{(1)} + 3\chi^{(3)}|\tilde{E}(\omega)|^2
	\label{eq:chieff}
\end{equation}

Also from the Kerr nonlinear optical relation, we may write d  $n=n_0 + \bar{n}+\left\langle \tilde{E}^2 \right\rangle $. Here,  $\bar{n_2}$ is a new
optical constant (the rate at which the refractive index increases with increasing optical intensity) than nonlinear refractive index ($n_2$). If we put the values $\tilde{E}$ from equation \ref{eq:Elcetrosynamics} in above expression, we can get $\left\langle \tilde{E}^2 \right\rangle$ = $2\tilde{E}(\omega)\tilde{E}(\omega)^*$ = $2\left| \tilde{E}(\omega)\right|^2 $. Hence, the  refractive index of materials can be expressed as,
\begin{equation}
	n = n_0 + 2 \bar{n_2} \left| \tilde{E}(\omega)\right|^2
	\label{eq:18}
\end{equation}

In order to relate the nonlinear susceptibility $\chi^{(3)}$ to the nonlinear refractive
index $n_2$, from the equation \ref{eq:chieff} and \ref{eq:18}, we can write it as,

\begin{equation}
	\begin{split}
		n^2  &= 1+ \chi_{eff}\\
		\left[ n_0 + 2 \bar{n_2} \left| \tilde{E}(\omega)\right|^2\right] ^2 &= 1+  \chi^{(1) }+ 3\chi^{(3)}\left| \tilde{E}(\omega)\right| ^2
		\end{split}
	\label{eq:n2}
\end{equation}

Expanding the term up to the order of  $\left| \tilde{E}(\omega)\right|^2 $, the above expression can be expanded as, $n^2_0 + 4n_0\bar{n_2}\left| \tilde{E}(\omega)\right|^2 ~=(1 + \chi^{(1)}) + [3\chi^{(3)}\left| \tilde{E}(\omega)\right|^2]$, which shows that the linear and nonlinear refractive indices are related to the linear and nonlinear susceptibilities by

\begin{equation}
	\begin{split}
	 n_0  &= \big(1+ \chi^{(1)}\big)^{1/2} \\
	& ~\& \\ 
	 \bar{n_2}  &= \frac{3\chi^{(3)}}{4n_0}
	\label{eq:n0}
\end{split}
\end{equation}

Also as we know the time-averaged intensity of the optical field which can be given by  $I = 2n_0\epsilon_0c\left| \tilde{E}(\omega)\right|^2$, where, from the Kerr equation we can get $2\bar{n_2}\left| \tilde{E}(\omega)\right|^2 = n_2I$. By solving the above expressions, we can get the relation between $n_2$ and $\bar{n_2}$ as follow:
\begin{equation}
	n_2 = \frac{\bar{n_2}}{n_0\epsilon_0c}
\end{equation}
Solving the $\bar{n_2}$, we can get $\chi^{(3)}$ as,
\begin{equation}
	\chi^3 = \frac{4n_0^2\epsilon_0c}{3}n_2 
\end{equation}
The above equation in SI and Gaussian unit can be express as,\cite{Shen1984, Boyd2020}

\begin{equation}
	\begin{split}
			n_2\left( \frac{m^2}{W}\right)  &=\frac{283}{n_0^2}\chi^{(3)}\left( \frac{m^2}{V^2}\right) ~~~~(SI)\\
			~~~
			n_2\left( \frac{cm^2}{W}\right)  &=\frac{12\pi^2}{n_0^2c}10^7\chi^{(3)}\left( \frac{m^2}{V^2}\right) ~~~~(esu)\\
	\end{split}
	\label{eq:n2final}
\end{equation}

\hrule

\newpage


\end{document}